\RequirePackage[l2tabu, orthodox]{nag}
\RequirePackage{ifpdf}
\RequirePackage[usenames,dvipsnames]{xcolor}
\RequirePackage{tikz}

\documentclass[prl,twocolumn,preprintnumbers,footinbib]{revtex4-1}%

\usepackage{hyperref}
\usepackage[utf8]{inputenc}
\usepackage{mathtools, amscd, dsfont, amsthm, amssymb, array}
\usepackage{graphicx}
\usepackage[toc,page]{appendix}
\usepackage{etoolbox}
\mathtoolsset{showonlyrefs=false}
\usetikzlibrary{arrows,calc,shapes.misc,decorations.markings,snakes}
\tikzset{cross/.style={cross out, draw=black, minimum size=2*(#1-\pgflinewidth), inner sep=0pt,
        outer sep=0pt},cross/.default={3pt},
    gluon/.style={decorate, decoration={coil,aspect=0.9,segment length=5pt, amplitude=3pt}}}

\usepackage[english]{babel}
\usepackage{epsf,epsfig}
\usepackage{amssymb,amsmath,amsfonts}
\usepackage{mathrsfs}
\usepackage{color}
\usepackage{enumitem}

\usepackage{tikz,subcaption}

\usepackage[mathscr]{eucal}

\DeclareCaptionJustification{rjustified}{\rjustified}
\makeatletter
\newcommand\rjustified{%
  \let\\\@fillcr
  \leftskip\z@\@plus -1fil
  \rightskip\z@\@plus 1fil
  \parfillskip\z@\@plus 0fil\relax
}
\makeatother

\def\xb{\bar{x}}

\def\tw{s}
\def\CB{\mathcal{B}}
\def\bF{\mathbf{F}}

\newcommand{\be}{\begin{equation}}
\newcommand{\ee}{\end{equation}}
\newcommand{\bea}{\begin{eqnarray}}
\newcommand{\eea}{\end{eqnarray}}

\newcommand{\nn}{\nonumber}

\begin{document}

\title{Bootstrapping string theory on AdS$_5 \times S^5$}

\author{J.~M.~Drummond, H.~Paul and M.~Santagata}
\affiliation{
$\rule{0pt}{.01cm}$
\makebox[\textwidth][c]{School of Physics and Astronomy, University of Southampton, Highfield, SO17 1BJ, United Kingdom}\\
}

\begin{abstract}
\noindent
We make an ansatz for the Mellin representation of the four-point amplitude of half-BPS operators of arbitrary charges at order $\lambda^{-\frac{5}{2}}$ in an expansion around the supergravity limit. Crossing symmetry and a set of constraints on the form of the spectrum uniquely fix the amplitude and double-trace anomalous dimensions at this order. The results exhibit a number of natural patterns which suggest that the bootstrap approach outlined here will extend to higher orders in a simple way.

\end{abstract}
\maketitle


\section{Introduction}

Recently great progress has been made in understanding the structure of amplitudes in anti-de-Sitter space by imposing consistency of the boundary conformal field theory. A particular case has been the focus of many investigations, namely $\mathcal{N}=4$ super Yang-Mills theory on the boundary which corresponds to type IIB superstrings interacting in the AdS${}_5\times S^5$ bulk \cite{Maldacena:1997re}. 

Physical quantities depend on the gauge coupling $g$ and the gauge group, which we take to be $SU(N)$. The holographic relation between the bulk and boundary theories implies that the spectrum of the conformal field theory is drastically simplified in the supergravity regime $ 0 \ll \lambda \ll N $ where $\lambda = g^2 N$ is the 't Hooft coupling. In this limit the spectrum is given by single-particle half-BPS operators and their multi-trace products while other operators, corresponding to excited string states, acquire infinite scaling dimensions in the limit and decouple.

We study four-point functions of single-particle operators in a double expansion in $1/N$ and $\lambda^{-\frac{1}{2}}$ around the supergravity limit. The leading large $N$ contributions to the operator product expansion (OPE) come from a degenerate spectrum of double-trace operators \cite{Alday:2017xua,Aprile:2017bgs,Aprile:2017xsp,Alday:2017vkk,Aprile:2017qoy,Aprile:2018efk,Alday:2018pdi,Alday:2018kkw,Aprile:2019rep,Alday:2019nin,Drummond:2019hel}.

The supergravity contribution to the four-point functions has a compact Mellin representation \cite{Rastelli:2016nze,Rastelli:2017udc}. The mixing between the double-trace operators can be resolved, yielding a very simple formula for the leading contributions to their anomalous dimensions \cite{Aprile:2017xsp,Aprile:2018efk}. In fact, the degeneracy is not fully lifted in supergravity. The residual partial degeneracy can be understood in terms of a surprising ten-dimensional conformal symmetry \cite{Caron-Huot:2018kta}.

Recent papers have explored the structure of string corrections to the tree-level supergravity amplitudes. Constraints from the flat-space limit \cite{Alday:2018pdi} and results derived using localisation \cite{Binder:2019jwn} allowed a family of correlation functions to be fixed at the first two non-trivial orders,  $\lambda^{-\frac{3}{2}}$ and $\lambda^{-\frac{5}{2}}$.
The order $\lambda^{-\frac{3}{2}}$ corrections can be determined for every half-BPS four-point function from the relevant term in the flat-space Virasoro-Shapiro amplitude \cite{Drummond:2019odu}. From these results it was found in \cite{Drummond:2019odu} that the double-trace spectrum reflected the ten-dimensional symmetry structure, even when taking into account the $\lambda^{-\frac{3}{2}}$ corrections. We explore this feature further here, generalising the results of previous papers to determine all half-BPS four-point functions up to order $\lambda^{-\frac{5}{2}}$.

We use an ansatz for the Mellin amplitude as a function of the external charges and minimal assumptions about the form of the corrections to the spectrum. Combined with crossing symmetry and OPE consistency the above is sufficient to determine the $\lambda^{-\frac{5}{2}}$ corrections to the correlation functions as well as the spectrum and three-point functions of the double-trace operators. The results reveal many beautiful features that are suggestive of a general pattern which should allow the method to be simply extended to yet higher orders in $\lambda^{-\frac{1}{2}}$. As observed in \cite{Drummond:2019odu}, we find that the ten-dimensional effective spin determines which operators receive string corrections to their dimensions and three-point functions. Moreover, at order $\lambda^{-\frac{5}{2}}$, we find that the partial degeneracy is broken at finite twist in a way consistent with other general features of the spectrum and suggestive of a general structure.


\section{Half-BPS four-point functions}
We recall that $\mathcal{N}=4$ super Yang-Mills theory has a spectrum of single-particle half-BPS operators given by
\be
\mathcal{O}_p(x,y) = y^{R_1}\ldots y^{R_p} {\rm tr}(\phi_{R_1} \ldots \phi_{R_p})(x) + \ldots\,.
\ee
Here $y^2=0$ and we omit $1/N$ suppressed multi-trace contributions determined by the condition that $\mathcal{O}_p$ should be orthogonal to all multi-trace operators \cite{Aprile:2018efk}.

Here we focus on four-point functions of such operators which, due to superconformal symmetry, have the form,
\be
\langle \mathcal{O}_{p_1} \mathcal{O}_{p_2} \mathcal{O}_{p_3} \mathcal{O}_{p_4} \rangle = \langle \mathcal{O}_{p_1} \mathcal{O}_{p_2} \mathcal{O}_{p_3} \mathcal{O}_{p_4} \rangle_{\rm free} + \mathcal{P}\, \mathcal{I} \, \mathcal{H}\,.
\label{partialNR}
\ee
The first term on the rhs is the contribution from free theory where $g=0$. The second term contains the factors $\mathcal{P}$ and $\mathcal{I}$, given in equations (\ref{calP}) and (\ref{calI}), and $\mathcal{H}(x,\bar{x};y,\bar{y})$ which encodes the dynamical contribution to the correlator. It depends on conformal and $su(4)$ cross-ratios,
\begin{align}
U &= x \bar{x} = \tfrac{x_{12}^2 x_{34}^2}{x_{13}^2 x_{24}^2}\,,  &&V = (1-x)(1-\bar{x}) = \tfrac{x_{14}^2 x_{23}^2}{x_{13}^2 x_{24}^2} \,, \notag \\
\tfrac{1}{\sigma} &= y \bar{y} = \tfrac{y_{12}^2 y_{34}^2}{y_{13}^2 y_{24}^2}\,,  &&\tfrac{\tau}{\sigma} = (1-y)(1-\bar{y}) = \tfrac{y_{14}^2 y_{23}^2}{y_{13}^2 y_{24}^2}\,.
\end{align}

Here we are concerned only with the leading large $N$ contribution to $\mathcal{H}$ corresponding to tree-level string amplitudes. This term admits an expansion in $\lambda^{-\frac{1}{2}}$,
\be
\mathcal{H} = \tfrac{1}{N^2}\bigl[\mathcal{H}^{(0)} + \lambda^{-\frac{3}{2}} \mathcal{H}^{(3)} + \lambda^{-\frac{5}{2}} \mathcal{H}^{(5)} + \ldots\bigr]\,.
\ee
The leading term $\mathcal{H}^{(0)}$ in the above expansion was determined for all external charges $\langle p_1 p_2 p_3 p_4 \rangle$ in \cite{Rastelli:2016nze,Rastelli:2017udc}, extending previous results (see e.g. \cite{Dolan:2006ec,Uruchurtu:2008kp,Uruchurtu:2011wh}) and verified by more recent supergravity analyses \cite{Arutyunov:2017dti,Arutyunov:2018neq,Arutyunov:2018tvn}.

As in \cite{Rastelli:2016nze,Rastelli:2017udc} we will use a Mellin representation,
\begin{align}
\mathcal{H}^{(n)}  &= \int \frac{ds}{2} \frac{dt}{2} U^\frac{s+p_3-p_4}{2} V^\frac{t-p_2-p_3}{2} \Gamma \mathcal{M}^{(n)}(s,t;\sigma,\tau)\,, \notag \\
\Gamma &= \Gamma \bigl[ \tfrac{p_1+p_2-s}{2}\bigr] \Gamma \bigl[\tfrac{p_3+p_4-s}{2} \bigr] \Gamma \bigl[ \tfrac{p_1+p_4-t}{2} \bigr] \notag \\
&\quad \,\, \Gamma \bigl[\tfrac{p_2+p_3-t}{2}\bigr] \Gamma \bigl[\tfrac{p_1+p_3-u}{2}\bigr] \Gamma \bigl[\tfrac{p_2+p_4-u}{2}\bigr]\,,
\label{Gamma}
\end{align}
where the Mandelstam-type variables $s,t,u$ obey
\be
s+t+u = 2\Sigma -4\,,\quad \Sigma=\tfrac{1}{2}(p_1 + p_2 + p_3 + p_4)\,.
\label{Mandrel}
\ee
We give the Mellin amplitude of \cite{Rastelli:2016nze,Rastelli:2017udc} in equation (\ref{M0}) in the Appendix. The most important point here is that it reduces to the flat-space supergravity amplitude in the large $s,t,u$ limit,
\be
\mathcal{M}^{(0)} \rightarrow B(\sigma,\tau)/(stu) \,, \quad B(\sigma,\tau) = {\textstyle \sum_{i,j}}\, \mathcal{N}_{ijk} \sigma^i \tau^j\,,
\ee
where the coefficients $\mathcal{N}_{ijk}$ are given in (\ref{Nijk}).
In fact, the large $s,t,u$ limit at each order in $\lambda^{-\frac{1}{2}}$ is controlled by the flat-space Virasoro-Shapiro amplitude $\mathcal{V}$ \cite{Penedones:2010ue,Goncalves:2014ffa,Alday:2018pdi,Binder:2019jwn},
\begin{align}\label{vs-flat}
\mathcal{V}&={\rm exp}\bigl\{ {\textstyle \sum}_{n\geq1} \tfrac{2\zeta_{2n+1} }{2n+1}(s^{2n+1}+t^{2n+1}+u^{2n+1})\bigr\}.
\end{align}
The precise relation between $\mathcal{V}$ and $\mathcal{M}$ requires an integral which gives the leading large $s,t,u$  behaviour:
\begin{align}
\label{flatlimitM3}
\mathcal{M}^{(3)} &\rightarrow  2^{-3}(\Sigma -1)_3 B(\sigma,\tau) \times 2\zeta_3\,, \\
\mathcal{M}^{(5)} & \rightarrow 2^{-5}(\Sigma -1)_5 B(\sigma,\tau) \times  \zeta_5 (s^2 + t^2 + u^2)\,.
\label{flatlimitM5}
\end{align}
The poles in the factor $\Gamma$ are due to unprotected double-trace operators exchanged in the OPE. The remaining poles in the supergravity Mellin amplitude $\mathcal{M}^{(0)}$ are  due to long single-trace contributions (or excited string state contributions) which must cancel against corresponding contributions present in the free-theory term in (\ref{partialNR}) since they should be absent from the supergravity spectrum. The $\lambda^{-\frac{1}{2}}$ corrections should then have no such poles and are therefore polynomial in $s,t,u$ \cite{Penedones:2010ue,Goncalves:2014ffa,Alday:2018pdi,Binder:2019jwn}. It follows \cite{Drummond:2019odu} that the result (\ref{flatlimitM3}) for $\mathcal{M}^{(3)}$ is in fact complete. The limit (\ref{flatlimitM5}) for $\mathcal{M}^{(5)}$, however, only determines the quadratic terms and does not specify additional linear and constant contributions in $s$ and $t$,
\begin{align}
\mathcal{M}^{(5)} &= \zeta_5 [2^{-5} (\Sigma -1)_5 B(\sigma,\tau) (s^2+t^2+u^2) \notag \\
&\quad + \alpha(\sigma,\tau) s + \beta(\sigma,\tau) t + \gamma(\sigma,\tau)]\,.
\end{align}
The coefficients $\alpha,\beta,\gamma$ are currently only known for external charges $\langle 22qq \rangle$ \cite{Alday:2018pdi,Binder:2019jwn} and, up to a single free parameter, $\langle 23\,q-1\,q\rangle$ \cite{Drummond:2019odu}, in which cases there is no dependence on $\sigma$ and $\tau$. In the case of $\langle 22qq \rangle$ we have
\begin{align}
&B = \tfrac{2^5 q^2}{(q-2)!}\,, \quad \alpha = \tfrac{(q)_5 2q^2(q-2)}{(q-2)!}\,, \quad \beta=0\,,\notag \\
& \gamma = -\tfrac{q (q)_4}{(q-2)!} 2(q^4+9q^3+10q^2-20q-25)\,.
\end{align}

To describe an ansatz for $\mathcal{M}^{(5)}$, it is helpful to parametrise the charges as
\be
\langle p_1 p_2 p_3 p_4 \rangle = \langle p-m \, p \,q-n \, q \rangle\,.
\ee
We use the $su(4)$ blocks $Y_{[aba]}(\sigma,\tau)$ (\ref{su4blocks}) instead of working with monomials in $\sigma$ and $\tau$,
\be
\label{alpha}
\alpha(\sigma,\tau) = (\Sigma-1)_4 {\textstyle \sum_{a,b}} B_{a,b} \tilde{\alpha}_{a,b} Y_{[aba]}(\sigma,\tau) \,,
\ee
and similarly for $\beta$, while for $\gamma$ we replace $(\Sigma-1)_4$ with $(\Sigma-1)_3$. We have included an explicit factor,
\be
B_{a,b} \!=\! \tfrac{pq(p-m)(q-n) (\Sigma-2)! b! (b+1)! (b+2+a)}{(p+r_1)!(p-r_2-2-a)!(q+r_3)!(q-r_4-2-a)! r_1 ! r_2 ! r_3 ! r_4 !}
\ee
where we use the notation
\be
r_1= \tfrac{b-m}{2}\,, \quad r_2 = \tfrac{b+m}{2}\, \quad r_3 = \tfrac{b-n}{2} \,, \quad r_4 = \tfrac{b+n}{2}\,.
\ee
The factor $B_{a,b}$ is in part motivated by the fact that
\be
B(\sigma,\tau) = 8 {\textstyle \sum_b}  B_{0,b} Y_{[0b0]}(\sigma,\tau)\,,
\ee
and also by the fact that for each $su(4)$ channel $[a,b,a]$ we can consistently make a polynomial ansatz for $\tilde{\alpha}$, $\tilde{\beta}$, $\tilde{\gamma}$ as a function of $p$ and $q$ for each required value of $m$ and $n$. 
Based on the observed structure of the $\langle 22qq \rangle$ amplitude, we allow $\tilde{\alpha}$ and $\tilde{\beta}$ to be quadratic and $\tilde{\gamma}$ to be quartic in $p$ and $q$.

Consistency with the $\langle 22 qq\rangle$ results and crossing symmetry imposes many constraints among the free parameters of the ansatz but cannot fix it uniquely.
To discuss the additional constraints we will impose, it is helpful to recall some facts about the double-trace spectrum.


\section{The double-trace spectrum}
At leading order in the large $N$ expansion, only double-trace multiplets are exchanged in the OPE. The primaries take the form
\be
\mathcal{O}_{pq} = \mathcal{O}_p \partial^l \Box^{\frac{1}{2}(\tau - p - q)} \mathcal{O}_q\,, \qquad ( p < q )\,.
\label{dbltrace}
\ee
For a given twist $\tau$, spin $l$ and $su(4)$ channel $[a,b,a]$, all the operators (\ref{dbltrace}) are degenerate at leading order in large $N$. We parametrise the unprotected ones as in \cite{Aprile:2018efk}
\begin{align}
p &= i +a + 1 + r\,, &&q=i+a+1+b-r\,, \notag \\
i &= 1,\ldots,(t-1)\,, &&r=0,\ldots,(\kappa -1)\,,
\end{align}
where we use the notation
\be\label{multiplicity}
t\equiv (\tau-b)/2-a\,,\quad 
\kappa \equiv   \left\{\begin{array}{ll}
\bigl\lfloor{\frac{b+2}2}\bigr\rfloor \quad &a+l \text{ even,}\\[.2cm]
\bigl\lfloor{\frac{b+1}2}\bigr\rfloor \quad &a+l \text{ odd.}
\end{array}\right.
\ee

For each $\vec{\tau}=(\tau,l,a,b)$, there are $d=\kappa(t-1)$ degenerate operators which mix, and we denote the range of values of $(p,q)$ by $\mathcal{D}_{\vec{\tau}}$. We will label the eigenstates $\mathcal{K}_{pq}$ with $p$ and $q$ parametrised by $i$ and $r$ as above. The mixing problem can be addressed  by considering the OPE. If we arrange a $(d \times d)$ matrix of correlators with the pairs $(p_1,p_2)$ and $(p_3,p_4)$ ranging over the same set $\mathcal{D}_{\vec{\tau}}$, we have
\begin{align}
&O( N^0): &&\langle \mathcal{O}_{p_1} \mathcal{O}_{p_2} \mathcal{O}_{p_3} \mathcal{O}_{p_4} \rangle_{\rm free}^{\rm long} = {\textstyle \sum_{\vec{\tau}}} A_{\vec{\tau}}\, \mathbb{L}_{\vec{\tau}}\,, \notag \\
&O(N^{-2}): &&  \qquad \qquad \,\, \mathcal{P} \, \mathcal{I} \, \mathcal{H}|_{\log u} = {\textstyle \sum_{\vec{\tau}}} M_{\vec{\tau}}  \mathbb{L}_{\vec{\tau}}\,,
\end{align}
where $A_{\vec{\tau}}$ and $M_{\vec{\tau}}$ are matrices of coefficients and $\mathbb{L}_{\vec{\tau}}$ is the superblock for long multiplets given in equation \eqref{longblocks} in the Appendix. The coefficients $A_{\vec{\tau}}$ are independent of $\lambda$ while $M_{\vec{\tau}}$ receives contributions at all orders where the corresponding $\mathcal{M}^{(n)}$ is non-zero.

The matrices $A_{\vec{\tau}}$ and $M_{\vec{\tau}}$ are related to three-point functions and anomalous dimensions of the $\mathcal{K}_{pq}$,
\begin{align}
\mathbb{C}_{\vec{\tau}}  \mathbb{C}_{\vec{\tau}} ^T = A_{\vec{\tau}} \,, \quad \mathbb{C}_{\vec{\tau}}  \eta_{\vec{\tau}}  \mathbb{C}_{\vec{\tau}}^T = M_{\vec{\tau}} \,.
\label{unmixing}
\end{align}
Here, $\mathbb{C}_{(pq),(\tilde{p}\tilde{q})}$ is a $(d\times d)$ matrix of three-point functions $[\langle \mathcal{O}_p \mathcal{O}_q \mathcal{K}_{\tilde{p}\tilde{q}} \rangle]$ and $\eta$ is a diagonal matrix encoding the anomalous dimensions of the eigenstates $\mathcal{K}_{pq}$:
\be
\Delta_{pq} = \tau-l + \tfrac{2}{N^2} \eta_{pq} + O(\tfrac{1}{N^4})\,.
\ee
The $\eta$ and $\mathbb{C}$ matrices are expanded for large $\lambda$ as,
\begin{align}
\eta_{pq} &= \eta^{(0)}_{pq} + \lambda^{-\frac{3}{2}} \eta^{(3)}_{pq} + \lambda^{-\frac{5}{2}} \eta^{(5)}_{pq} + \ldots\, ,\notag  \\
\mathbb{C} &= \mathbb{C}^{(0)} + \lambda^{-\frac{3}{2}} \mathbb{C}^{(3)} + \lambda^{-\frac{5}{2}} \mathbb{C}^{(5)} + \ldots\,.
\end{align}

The tree-level contributions $\eta^{(0)}$ induced by (\ref{M0}) take an astonishingly simple form \cite{Aprile:2018efk},
\be
\eta^{(0)}_{pq} = - 2 M_t M_{t+l+1}/(\ell_{10}+1)_6\,,
\ee
where the numerator is given by
\be
M_t = (t-1)(t+a)(t+a+b+1)(t+2a+b+2),
\ee
and the denominator is a Pochhammer of the effective ten-dimensional spin
\be\label{l10}
\ell_{10}(p) = l + a +2(i+r) -1 - \tfrac{1+(-1)^{a+l}}{2}\,.
\ee
In \cite{Caron-Huot:2018kta} it was recognised that the appearance of $\ell_{10}$ signals the presence of a ten-dimensional conformal symmetry. Note that $\ell_{10}$ only depends on the combination $i+r$ (or $p$, not $q$), so in general there are several states with the same anomalous dimension, and the resolution of the operator mixing in tree-level supergravity is only partial \cite{Aprile:2018efk}, as depicted in Figure \ref{fig:spectrum_sugra}. This means that, although the eigenvalue problem is well posed, the leading-order three-point functions $\mathbb{C}^{(0)}$ are in general not fully determined by $\mathcal{M}^{(0)}$.

\begin{figure}
\begin{center}
\begin{tikzpicture}[scale=0.3]
\definecolor{lightgray}{gray}{0.87}
\definecolor{darkgray}{gray}{0.3}
\def\zero{(0,0)}
\def\pta{(2,8)}
\def\ptb{(5,5)}
\def\ptc{(11,11)}
\def\ptd{(8,14)}
\def\newzero{(0,2.5)}
\draw[thick,->] \newzero -- (2.5,2.5) node[anchor=west] {$p$};
\draw[thick,->] \newzero -- (0,5) node[anchor=south] {$q$};
\draw[darkgray,dashed,-latex] (1,7) -- (9.5,15.5) node[anchor=north west] {$i$};
\draw[darkgray,dashed,-latex] (1,9) -- (6.5,3.5) node[anchor=south west] {$r$};
\draw \pta -- \ptb;
\draw \pta -- \ptd;
\draw \ptb -- \ptc;
\draw \ptc -- \ptd;
\foreach \r in {0,1,2,3}
\foreach \i in {1,...,7}
\filldraw (\i+\r+1,\i-\r+7) circle (0.18);
\node[scale=0.5,label=left:$A$] at \pta {};
\node[scale=0.5,label=below:$B$] (a) at \ptb {};
\node[scale=0.5,label=right:$C$] (a) at \ptc {};
\node[scale=0.5,label=above:$D$] (a) at \ptd {};
\draw[line width=0.3pt] (3,9) -- (3,7);
\draw[line width=0.3pt] (4,10) -- (4,6);
\draw[line width=0.3pt] (5,11) -- (5,5);
\draw[line width=0.3pt] (6,12) -- (6,6);
\draw[line width=0.3pt] (7,13) -- (7,7);
\draw[line width=0.3pt] (8,14) -- (8,8);
\draw[line width=0.3pt] (9,13) -- (9,9);
\draw[line width=0.3pt] (10,12) -- (10,10);
\node[scale=0.9] (legend) at (20,5) {$\begin{array}{l}  
										\displaystyle A=(a+2,a+b+2) \\[.1cm]
										\displaystyle B=(a+\mu+1,a+b-\mu+3) \\[.1cm]
										\displaystyle C=(a+t+\mu-1,a+b+t-\mu+1) \\[.1cm]
										\displaystyle D=(a+t,a+b+t) \\[.1 cm] \end{array}$};
\end{tikzpicture}
\end{center}\vspace{-0.57cm}\captionsetup{format=plain,justification=rjustified,labelsep=period}\caption{Spectrum of anomalous dimensions $\eta^{(0)}_{pq}$ of the double-trace eigenstates $\mathcal{K}_{pq}$, represented by dots in the $(p,q)$-plane. Anomalous dimensions which remain degenerate are connected by vertical lines of constant $p$.}\label{fig:spectrum_sugra}
\end{figure}

The first string corrections $\eta^{(3)}$ are even simpler \cite{Drummond:2019odu}. They are only non-zero for $l=a=0$ and $i=1$, $r=0$, (or $\ell_{10} = 0$) where there is no partial degeneracy in the supergravity spectrum, corresponding to the leftmost corner labelled by $A$ in Figure \ref{fig:spectrum_sugra}. They take the form
\be
\eta^{(3)} = -\tfrac{1}{840} M_t M_{t+l+1} \zeta_3 (t-1)_3 (t+b+1)_3 \,.
\ee
Note that $\eta^{(0)}$ is a factor and the total polynomial degree in $t$ is 14. The fact that $\eta^{(3)}$ depends only on $\ell_{10}$, instead of $l,a,i,r$ individually, suggests that the ten-dimensional conformal symmetry is respected also at order $\lambda^{-\frac{3}{2}}$. The corrections to the three-point functions are uniquely determined and vanish, $\mathbb{C}^{(3)}=0$.

The above result generalises simply to states of the highest possible spin $l=(n-3)$ at order $\lambda^{-\frac{n}{2}}$ with $n$ odd. In this case the only relevant terms in $\mathcal{M}^{(n)}$ are the highest powers in $s,t,u$ which are determined by the flat-space limit. Again these terms are only non-zero for $a=0$ and $i=1$, $r=0$ and we find that $\mathbb{C}^{(n)}|_{l=n-3}=0$ with the anomalous dimension given by
\be
\eta^{(n)}_{l=n-3} \propto -M_t M_{t+l+1} \zeta_n (t-1)_n (t+b+1)_n \,.
\label{maxspinanomdim}
\ee
Note the anomalous dimensions are invariant under
\be
t \mapsto - t -b -2a - l - 2 \,.
\label{symmetry}
\ee

As argued in \cite{Drummond:2019odu}, the ten-dimensional conformal symmetry, present in the supergravity anomalous dimensions $\eta^{(0)}_{pq}$, assigns an effective ten-dimensional spin $\ell_{10}$ to each eigenstate $\mathcal{K}_{pq}$ by means of equation \eqref{l10}. The above result \eqref{maxspinanomdim} then suggests that this assignment is respected by the (tree-level) string corrections, to any order in $\lambda^{-\frac{1}{2}}$, i.e. the maximal exchanged spin $\ell_{10}$ at a given order $\lambda^{-\frac{n}{2}}$ in the flat-space Virasoro-Shapiro amplitude \eqref{vs-flat} determines which eigenstates $\mathcal{K}_{pq}$ develop an anomalous dimension, as well as which three-point functions are non-zero. We emphasise that this does not imply that the conformal symmetry is preserved by the string corrections -- on the contrary, it will turn out that the $\lambda^{-\frac{5}{2}}$ corrections actually break it, albeit in a way consistent with the assignment \eqref{l10}.

Based on the above observations, we propose the following conditions on the double-trace data at order $\lambda^{-\frac{n}{2}}$:
\begin{align}
\label{eta10dspin}
\bullet \,\,&\eta^{(n)}_{pq} = 0 \text{ for } \ell_{10}(p) > n-3\,, \\
\label{C10dspin}
\bullet \,\,&\mathbb{C}^{(n)}_{(pq),(\tilde{p}\tilde{q})} = 0 \text{ for } \ell_{10}(\tilde{p})>n-3\,, \\
\label{etapoly}
\bullet \,\,&\eta^{(n)}_{i=1,r=0} \text{ is polynomial in $t$ of degree $8+2n$}\,, \\
\label{etalimit}
\bullet \,\,&\eta^{(n)}_{pq} \text{ only depends on } \ell_{10}(p) \text{ as } t\rightarrow \infty\,.
\end{align}
The constraint (\ref{eta10dspin}) says that $\ell_{10}$ dictates the non-zero contributions to $\eta$ and generalises the highest-spin $l=n-3$, $a=0$ result from equation \eqref{maxspinanomdim}. Similarly, the condition (\ref{C10dspin}) says that the columns of $\mathbb{C}^{(n)}$ corresponding to operators with too high ten-dimensional spin vanish. In the $n=3$ case it implies $\mathbb{C}^{(3)}=0$, since the first equation in (\ref{unmixing}) implies up to rescaling that $\mathbb{C}^{(0)}$ is an orthogonal matrix. Using the fact that $A_{\vec{\tau}}$ is independent of $\lambda$, its first correction $\mathbb{C}^{(3)}$ obeys
\be
\mathbb{C}^{(3)} \mathbb{C}^{(0) T} + \mathbb{C}^{(0)} \mathbb{C}^{(3) T} = 0\,,
\ee
and therefore, after a change of basis, it is antisymmetric. If all but the first column vanishes then the whole matrix vanishes. Importantly, for $n=5$ the same condition is weaker than the condition $\mathbb{C}^{(5)}=0$ examined in \cite{Drummond:2019odu} since now there are generically three non-zero columns. 

The condition (\ref{etapoly}) is an assumption on the anomalous dimension in the case of no partial degeneracy. The polynomial should obey the symmetry (\ref{symmetry}) and is of the same order as in the maximal spin case (\ref{maxspinanomdim}). The fourth condition (\ref{etalimit}) was also observed in \cite{Drummond:2019odu}, albeit under the (erroneously) stronger assumption $\mathbb{C}^{(5)}=0$. It relates to the restoration of ten-dimensional Lorentz symmetry in the flat-space limit (corresponding to $t\rightarrow \infty$).


\section{Results}
Imposing the conditions (\ref{eta10dspin})--(\ref{etalimit}) in the case $n=5$, we find a unique consistent solution for the Mellin amplitude and the spectrum. We emphasise that the existence of a solution consistent with the ansatz for the Mellin amplitude, crossing symmetry and the spectrum constraints is highly non-trivial. Actually, various computations in some channels have revealed that the constraints \eqref{eta10dspin} and \eqref{C10dspin} are really a consequence of imposing \eqref{etapoly} and an ansatz of the form \eqref{alpha} for the Mellin amplitude.

Here we summarise the form of the $\lambda^{-\frac{5}{2}}$ amplitude and the spectrum resulting from the above assumptions. First, we find that the $su(4)$ channels are constrained by $a \leq 2$, consistent with the ten-dimensional spin obeying $\ell_{10} \leq 2$ at this order. The resulting partial wave coefficients are
\begin{align}
\tilde{\alpha}_{2,b} &= \tilde{\beta}_{2,b} = 0\,, \quad \tfrac{1}{2} \tilde{\gamma}_{2,b} = -\tilde{\alpha}_{1,b} = -\tfrac{1}{2}\tilde{\beta}_{1,b} = 1\,, \notag \\
\tilde{\gamma}_{1,b} &= 2 ( \tfrac{ m n}{ 4 b_1} (\tilde{p}\tilde{q}+ b_1)+ ( \Sigma^2-4 ) )\,, \notag \\
\tilde{\alpha}_{0,b} &=-\tfrac{1}{8} (3+ \tfrac{m n}{ b_0} ) (\tilde{p}\tilde{q}+b_0)+\tfrac{1}{2}( \Sigma^2 - 4)\,, \notag \\
\tilde{\beta}_{0,b} &= -\tfrac{n m}{4 b_0} (\tilde{p} \tilde{q} + b_0)\,, \notag \\
\tilde{\gamma}_{0,b} &= -\tfrac{1}{128}\bigl[\tfrac{A}{b_0-5} + \tfrac{B}{b_0} + C\bigr]\,.
\end{align}
Here we define $\tilde{p} =(2p-m)$, $\tilde{q}=(2q-n)$ and $b_a=b(b+4+2a)$, while for $\tilde{\gamma}_{0,b}$ we have (using $R=\tilde{p}\tilde{q}+b_0+8$)
\begin{align}
A &= -(m^2-1)(n^2-1)(\tilde{p}^2-1)(\tilde{q}^2-1)\,, \notag \\
B &= m n \tilde{p}\tilde{q} [m n (\tilde{p}\tilde{q} - 8) -32 (\Sigma^2 - 4)]\,,\notag \\
C &= 16(\Sigma^2-4)((2\Sigma+1)^2-2mn) -5 m^2 n^2 +195   \notag \\
&+4b_0(\Sigma+4)^2+4\Sigma(9\Sigma-8R)-13R^2 -74R+177b_0 \notag \\
&+ (m^2+n^2)(2R-4(\Sigma-2)^2-b_0+141) \,.
\end{align}

The anomalous dimensions are non-vanishing only for $\ell_{10}\leq 2$, constraining the possible values of $(i,r,l,a)$. To write the anomalous dimensions $\eta^{(5)}_{i,r|l,a}$, we define the polynomial $\mathcal{T}$ as follows:
\begin{align}
N_t &= (t-1) (t+a) (t+a+b+1)\,, \notag \\
 \mathcal{T}_{t,l,a,b} &=\tfrac{1}{166320}\zeta_5 M_t M_{t+l+1} N_tN_{-t-2a-b-l-2} \,.
\end{align}
Note that $ \mathcal{T}_{t,0,0,b} \propto \eta^{(3)}(t,b)$.
For spin two we must have $i=1$, $r=0$, $a=0$ and we find
\be
{\eta^{(5)}_{1,0|2,0}} = \mathcal{T}_{t,2,0,b}(t+1) (t+2) (t+b+2) (t+b+3),
\ee
which is just a particular case of equation \eqref{maxspinanomdim}.

For spin one we have $i=1$, $r=0$ and $a=0,1$:
\begin{align}
{\eta^{(5)}_{1,0|1,0}} &= \tfrac{1}{2}\mathcal{T}_{t,1,0,b} (t+1) (t+b+2) ( 2 t (3 + b + t)+b)\,, \notag \\
{\eta^{(5)}_{1,0|1,1}} &= \mathcal{T}_{t,1,1,b} t (t+2) (t+b+3) (t+b+5)\,.
\end{align}
The spin-zero anomalous dimensions have support on $a=0,1,2$. For $a=1,2$ we have only $i=1$, $r=0$,
\begin{align}
{\eta^{(5)}_{1,0|0,1}} &= \tfrac{1}{2}\mathcal{T}_{t,0,1,b}t (t+b+4) ( 2 t^2 +  2(4 + b)t+b+6)\,, \notag \\
{\eta^{(5)}_{1,0|0,2}} &=  \mathcal{T}_{t,0,2,b}t (1 + t) (5 + b + t) (6 + b + t)\,.
\end{align}
In all the above cases we have $\mathbb{C}^{(5)}=0$. A pictorial representation of the spectrum for those cases is given in Figure \ref{fig:a}. On the other hand, the case $a=0$ allows for generically three non-zero components, depending on the values of $t$ and $b$. Using $\theta\equiv \tau+2=2t+2+b$, the $i=1$ component reads
\begin{align}
{\eta^{(5)}_{1,0|0,0}} &=  \tfrac{77}{18} \mathcal{T}_{t,0,0,b}\,f_{b,t}\,, \notag \\
f_{b,t} &= \tfrac{9}{4}(\theta^2-b_0)^2-35(\theta^2-b_0)-34b_0+639\,.
\end{align}
Finally, the $(i,r)=(1,1)$ and $(2,0)$ components read
\begin{align}\label{degenlift}
{\eta^{(5)}_{2,0|0,0}} &=\tfrac{1}{9}\mathcal{T}_{t,0,0,b}  \bigl(j_{b, t} - 10 \sqrt{k_{b, t}}\bigr)\,, \notag \\
{\eta^{(5)}_{1,1|0,0}} &=\tfrac{1}{9}\mathcal{T}_{t,0,0,b}  \bigl(j_{b, t} + 10 \sqrt{k_{b, t}}\bigr)\,, \notag \\
j_{b,t} &= \tfrac{1}{4}f_{b,t} -\tfrac{15}{4}(\theta^2+b_0+21) \, , \notag\\ 
k_{b,t} &=  j_{b,t}+(\theta^2+b_0)(\theta^2+b_0-10)\,.
\end{align}
Note that the residual partial degeneracy is lifted by the square root, as shown in Figure \ref{fig:b}. Moreover, in the $l=a=0$ case, we have $\mathbb{C}^{(5)} \neq 0$.

\begin{figure}
\begin{center}
\begin{subfigure}[b]{0.2\textwidth}
\centering
\begin{tikzpicture}[scale=0.3]
\definecolor{lightgray}{gray}{0.87}
\definecolor{darkgray}{gray}{0.3}
\def\zero{(0,0)}
\def\pta{(2,8)}
\def\ptb{(5,5)}
\def\ptc{(12,12)}
\def\ptd{(9,15)}
\def\ptc{(11,11)}
\def\ptd{(8,14)}
\def\newzero{(0,2)}
\draw \pta -- \ptb;
\draw \pta -- \ptd;
\draw \ptb -- \ptc;
\draw \ptc -- \ptd;
\draw[darkgray,dashed,-latex] (1,7) -- (9.5,15.5) node[anchor=north west] {$i$};
\draw[,darkgray,dashed,-latex] (1,9) -- (6.5,3.5) node[anchor=south west] {$r$};
\foreach \r in {0,1,2,3}
\foreach \i in {1,...,7}
\filldraw (\i+\r+1,\i-\r+7) circle (0.07);
\filldraw (2,8) circle (0.18);
\node[scale=0.5,label=left:$A$] at \pta {};
\node[scale=0.5,label=below:$B$] (a) at \ptb {};
\node[scale=0.5,label=right:$C$] (a) at \ptc {};
\node[scale=0.5,label=above:$D$] (a) at \ptd {};
\end{tikzpicture}\caption{}\label{fig:a}
\end{subfigure}\hspace{0.5cm}
\begin{subfigure}[b]{0.2\textwidth}
\centering
\begin{tikzpicture}[scale=0.3]
\definecolor{lightgray}{gray}{0.87}
\definecolor{darkgray}{gray}{0.3}
\def\zero{(0,0)}
\def\pta{(2,8)}
\def\ptb{(5,5)}
\def\ptc{(12,12)}
\def\ptd{(9,15)}
\def\ptc{(11,11)}
\def\ptd{(8,14)}
\def\newzero{(0,2)}
\draw \pta -- \ptb;
\draw \pta -- \ptd;
\draw \ptb -- \ptc;
\draw \ptc -- \ptd;
\draw[darkgray,dashed,-latex] (1,7) -- (9.5,15.5) node[anchor=north west] {$i$};
\draw[,darkgray,dashed,-latex] (1,9) -- (6.5,3.5) node[anchor=south west] {$r$};
\foreach \r in {0,1,2,3}
\foreach \i in {1,...,7}
\filldraw (\i+\r+1,\i-\r+7) circle (0.07);
\filldraw (2,8) circle (0.18);
\filldraw (3,7) circle (0.18);
\filldraw (3,9) circle (0.18);
\node[scale=0.5,label=left:$A$] at \pta {};
\node[scale=0.5,label=below:$B$] (a) at \ptb {};
\node[scale=0.5,label=right:$C$] (a) at \ptc {};
\node[scale=0.5,label=above:$D$] (a) at \ptd {};
\draw [to-to, line width=0.5pt] (3,7.3) -- (3,8.7);
\end{tikzpicture}\caption{}\label{fig:b}
\end{subfigure}
\end{center}\vspace{-0.6cm}\captionsetup{format=plain,justification=rjustified,labelsep=period}\caption{Depiction of anomalous dimensions $\eta^{(5)}_{i,r|l,a}$: the non-vanishing ones are denoted by filled circles, all others are zero. Diagram (a) describes the cases $(l,a)=(2,0)$, $(1,1)$, $(1,0)$, $(0,2)$, $(0,1)$, where only one anomalous dimension is non-zero. Diagram (b) shows the case $(l,a)=(0,0)$, where the arrow indicates the lifting of the residual degeneracy for $(i,r)=(1,1)$ and $(2,0)$.}\label{fig:spectrum_corrected}
\end{figure}


\section{Discussion and outlook}
The results of the previous section provide a Mellin formula for all correlators at order $\lambda^{-\frac{5}{2}}$, as well as the corrections to the spectrum. The correlators are consistent with the results for $\langle 22 qq \rangle$ \cite{Alday:2018pdi,Binder:2019jwn} given above and $\langle 23 \,q-1\, q\rangle$ derived in \cite{Drummond:2019odu}. Note that the anomalous dimensions found here differ from those conjectured in \cite{Drummond:2019odu} since we have found here that $\mathbb{C}^{(5)}\neq 0$ in general.

In the first case, where residual degeneracy is present in the supergravity spectrum, the $\lambda^{-\frac{5}{2}}$ corrections resolve it. Due to the residual two-fold mixing problem, the appearance of square roots in the anomalous dimension is to be expected; this did not happen in supergravity due to the ten-dimensional conformal symmetry. In some cases the square roots in (\ref{degenlift}) have to disappear:
\begin{itemize}
\item When $t=2$, there is no degeneracy and only two states acquire anomalous dimension. In fact, $k_{b,2}= j_{b,2}^2/100$ and ${\eta^{(5)}_{1,1|0,0}}$ becomes a rational function. 
\item When $b=0$, $b=1$, there is no degeneracy for any $t$ ($ \kappa=1$ in \eqref{multiplicity}): the square roots disappear again.
\item In the flat-space limit $t\rightarrow \infty$, the square-root terms are suppressed and degeneracy is restored, respecting the ten-dimensional Lorentz symmetry.
\end{itemize}
The disappearance of the square roots in these cases is a strong check of the consistency of the solution. Finally, all the anomalous dimensions have some shared features:
\begin{itemize}
\item When expressed in terms of the twist $\tau$ (or $\theta=2t+2a+b+l+2$) instead of $t$, they really depend on the $su(4)$ labels only through the Casimir combination $b_a= b(b+4+2a)$.
\item They enjoy the supergravity symmetry \eqref{symmetry}: this in turn means that all the quartic polynomials $f,j,k$ are actually quadratic in $\theta^2$. We partly imposed this property in (\ref{etapoly}), but again in many examples it was found to follow from the other assumptions.
\end{itemize}

As mentioned earlier, the bootstrap constraints \eqref{eta10dspin}--\eqref{etalimit} were motivated by the result \eqref{maxspinanomdim} for the highest-spin anomalous dimension and its agreement with the assignment of the ten-dimensional spin according to \eqref{l10}. The former statement is valid at any order $\lambda^{-\frac{n}{2}}$, with $n$ odd. We thus believe that the methods developed here will continue to be effective at higher orders in $\lambda^{-\frac{1}{2}}$, the next case being $\lambda^{-3}$. It will be interesting to examine the first case of triple residual degeneracy at order $\lambda^{-\frac{7}{2}}$ to see if there is hope for an explicit formula for the spectrum. We hope that this may allow us to apply a bootstrap approach to the full classical string amplitude in AdS.

This in turn will provide valuable information on the $\lambda$ dependence of the loop amplitudes.
In fact, considering correlators of generic external charges at loop order and studying their mutual consistency under crossing might provide a way to further substantiate the validity of our bootstrap method.


\section*{Acknowledgements}
We thank Dhritiman Nandan and Kostas Rigatos for collaboration on related topics and Francesco Aprile, Davide Bufalini, Paul Heslop, and Sami Rawash for interesting discussions. This work was supported in part by ERC Consolidator Grant No. 648630 IQFT.


\appendix\setcounter{equation}{0}\renewcommand{\theequation}{A\arabic{equation}}
\section{Appendix: Supergravity Mellin amplitude and superconformal blocks}\label{App-RZ}
We give here the Mellin amplitude in supergravity,
\begin{align}
\mathcal{M}^{(0)} = \sum_{i,j} \frac{\mathcal{N}_{ijk} \sigma^i \tau^j}{(s-\tilde{s}+2k)(t-\tilde{t}+2j)(u-\tilde{u}+2i)}\,,
\label{M0}
\end{align}
with $i+j+k = p_3 + {\rm min}(0,\frac{p_{13}+p_{24}}{2}) -2$ and the sum taken such that $i,j,k\geq0$. Here we have used
\begin{align}
\tilde{s} &= {\rm min}(p_1+p_2,p_3+p_4)-2\,, \notag \\ \tilde{t} &= p_2+p_3-2\,, \quad \tilde{u} = p_1+p_3 - 2\,.
\end{align}
Finally, the coefficients $\mathcal{N}_{ijk}$ are given by \cite{Rastelli:2016nze,Drummond:2019odu}
\be
\mathcal{N}_{ijk} = \frac{8 p_1 p_2 p_3p _4 (i!j!k!)^{-1}}{\bigl[ \frac{p_{43}+p_{21}+2i}{2}\bigr]! \bigl[ \frac{p_{43}-p_{21}+2j}{2}\bigr]! \bigl[ \frac{|p_{13}+p_{24}|+2k}{2}\bigr]!}.
\label{Nijk}
\ee

The relevant superblocks for long multiplets were given in \cite{Dolan:2004iy,Doobary:2015gia}. In our notation, they take the form
\be
\mathbb{L}_{\vec{\tau}} = \mathcal{P} \, \mathcal{I} \, u^{\frac{p_{34}}{2}-2} Y_{[aba]}(y,\bar{y}) \mathcal{B}^{2+\frac{\tau}{2}|l}(x,\bar{x})\,.
\label{longblocks}
\ee
In (\ref{longblocks}) we have
\be
\mathcal{P} = N^{\frac{1}{2}\sum p_i } g_{12}^{\frac{p_1+p_2-p_{43} }{2} } \,g_{14}^{\frac{-p_{21}+p_{43}}{2}}\,  g_{24}^{\frac{p_{21}+p_{43}}{2} } g_{34}^{{p_3}{} }\, ,
\label{calP}
\ee
where we introduce the propagators $g_{ij}$:
\be
g_{ij} = y_{ij}^2 / x_{ij}^2\,, \quad x_{ij}^2 = (x_i-x_j)^2,\quad y_{ij}^2 = y_i \cdot y_j\,.
\ee
The factor $\mathcal{I}$ in (\ref{longblocks}) is given by
\be
\mathcal{I}(x,\bar{x};y,\bar{y}) = (x-y)(x-\bar{y})(\bar{x}-y)(\bar{x}-\bar{y})/(y \bar{y})^2 \,,
\label{calI}
\ee
and is present due to superconformal symmetry \cite{Eden:2000bk,Dolan:2004iy}.
The $su(4)$ blocks for $[a,b,a] = [\mu-\nu,2\nu+p_{43},\mu-\nu]$ are given in terms of Jacobi polynomials $J_\mu^{(\alpha,\beta)}$:
\begin{align}
\label{su4blocks}
&Y_{[aba]}(y,\bar y) = (P_\nu(y)P_{\mu+1}(\bar{y})-P_{\mu+1}(y)P_\nu(\bar{y}))/(y-\bar{y})\,, \notag \\
&P_\mu(y) = \tfrac{\mu ! y}{(\mu+1+p_{43})_\mu} J_\mu^{\bigl(\tfrac{p_{43}-p_{21}}{2},\tfrac{p_{21}+p_{43}}{2}\bigr)}\bigl(\tfrac{2}{y}-1\bigr)\,.
\end{align}
Finally, the conformal blocks are given by
\bea
\label{Confblock}
\CB^{\,\tw|l}(x,\xb)&=&(-1)^l\  \tfrac{u^\tw x^{l+1}\, \bF_{\tw+l
}(x)\bF_{\tw-1}(\xb)- (x\leftrightarrow \xb)  }{x-\xb}, \nn\\
\bF_{\tw}(x)&=&~_2F_1\big({\tw- \tfrac{p_{12}}{2},
\tw+\tfrac{p_{34}}{2}; 2\tw};x\big)\,.
\eea

\bibliographystyle{apsrev4-1}

\begin{thebibliography}{99}

\bibitem{Maldacena:1997re}
  J.~M.~Maldacena,
  Int.\ J.\ Theor.\ Phys.\  {\bf 38} (1999) 1113
   [Adv.\ Theor.\ Math.\ Phys.\  {\bf 2} (1998) 231]
  [hep-th/9711200].

\bibitem{Alday:2017xua}
  L.~F.~Alday and A.~Bissi,
  Phys.\ Rev.\ Lett.\  {\bf 119} (2017) no.17,  171601
  [arXiv:1706.02388 [hep-th]].

\bibitem{Aprile:2017bgs}
  F.~Aprile, J.~M.~Drummond, P.~Heslop and H.~Paul,
  JHEP {\bf 1801} (2018) 035
  [arXiv:1706.02822 [hep-th]].

\bibitem{Aprile:2017xsp}
  F.~Aprile, J.~M.~Drummond, P.~Heslop and H.~Paul,
  JHEP {\bf 1802} (2018) 133
  [arXiv:1706.08456 [hep-th]].

\bibitem{Alday:2017vkk}
  L.~F.~Alday and S.~Caron-Huot,
  JHEP {\bf 1812} (2018) 017
  [arXiv:1711.02031 [hep-th]].

\bibitem{Aprile:2017qoy}
  F.~Aprile, J.~M.~Drummond, P.~Heslop and H.~Paul,
  JHEP {\bf 1805} (2018) 056
  [arXiv:1711.03903 [hep-th]].

\bibitem{Aprile:2018efk}
  F.~Aprile, J.~Drummond, P.~Heslop and H.~Paul,
  Phys.\ Rev.\ D {\bf 98} (2018) no.12,  126008
  [arXiv:1802.06889 [hep-th]].

\bibitem{Alday:2018pdi}
  L.~F.~Alday, A.~Bissi and E.~Perlmutter,
  JHEP {\bf 1906} (2019) 010
  [arXiv:1809.10670 [hep-th]].

\bibitem{Alday:2018kkw}
  L.~F.~Alday,
  JHEP \textbf{04} (2021), 005
  [arXiv:1812.11783 [hep-th]].

  \bibitem{Aprile:2019rep}
  F.~Aprile, J.~Drummond, P.~Heslop and H.~Paul,
  JHEP \textbf{03} (2020), 190
  [arXiv:1912.01047 [hep-th]].

  \bibitem{Alday:2019nin}
  L.~F.~Alday and X.~Zhou,
  JHEP \textbf{09} (2020), 008
  [arXiv:1912.02663 [hep-th]].

  \bibitem{Drummond:2019hel}
  J.~M.~Drummond and H.~Paul,
  JHEP \textbf{03} (2021), 038
  [arXiv:1912.07632 [hep-th]].
  
\bibitem{Rastelli:2016nze}
  L.~Rastelli and X.~Zhou,
  Phys.\ Rev.\ Lett.\  {\bf 118} (2017) no.9,  091602
  [arXiv:1608.06624 [hep-th]].
  
\bibitem{Rastelli:2017udc}
  L.~Rastelli and X.~Zhou,
  JHEP {\bf 1804} (2018) 014
  [arXiv:1710.05923 [hep-th]].
  
\bibitem{Caron-Huot:2018kta}
  S.~Caron-Huot and A.~K.~Trinh,
  JHEP {\bf 1901} (2019) 196
  [arXiv:1809.09173 [hep-th]].
  
\bibitem{Binder:2019jwn}
  D.~J.~Binder, S.~M.~Chester, S.~S.~Pufu and Y.~Wang,
  JHEP {\bf 1912} (2019) 119
  [arXiv:1902.06263 [hep-th]].

\bibitem{Drummond:2019odu}
  J.~M.~Drummond, D.~Nandan, H.~Paul and K.~S.~Rigatos,
  JHEP {\bf 1912} (2019) 173
  [arXiv:1907.00992 [hep-th]].

\bibitem{Dolan:2006ec}
  F.~A.~Dolan, M.~Nirschl and H.~Osborn,
  Nucl.\ Phys.\ B {\bf 749} (2006) 109
  [hep-th/0601148].

\bibitem{Uruchurtu:2008kp}
  L.~I.~Uruchurtu,
  JHEP {\bf 0903} (2009) 133
  [arXiv:0811.2320 [hep-th]].
  
\bibitem{Uruchurtu:2011wh}
  L.~I.~Uruchurtu,
  JHEP {\bf 1108} (2011) 133
  [arXiv:1106.0630 [hep-th]].

\bibitem{Arutyunov:2017dti} 
  G.~Arutyunov, S.~Frolov, R.~Klabbers and S.~Savin,
  JHEP {\bf 1704}, 005 (2017)
  [arXiv:1701.00998 [hep-th]].

\bibitem{Arutyunov:2018neq}
  G.~Arutyunov, R.~Klabbers and S.~Savin,
  JHEP {\bf 1809} (2018) 023
  [arXiv:1806.09200 [hep-th]].

\bibitem{Arutyunov:2018tvn}
  G.~Arutyunov, R.~Klabbers and S.~Savin,
  JHEP {\bf 1809} (2018) 118
  [arXiv:1808.06788 [hep-th]].
  
\bibitem{Penedones:2010ue}
  J.~Penedones,
  JHEP {\bf 1103} (2011) 025
  [arXiv:1011.1485 [hep-th]].

\bibitem{Goncalves:2014ffa}
  V.~Gonçalves,
  JHEP {\bf 1504} (2015) 150
  [arXiv:1411.1675 [hep-th]].

\bibitem{Dolan:2004iy}
  F.~A.~Dolan and H.~Osborn,
  Annals Phys.\  {\bf 321} (2006) 581
  [hep-th/0412335].

\bibitem{Doobary:2015gia}
  R.~Doobary and P.~Heslop,
  JHEP {\bf 1512} (2015) 159
  [arXiv:1508.03611 [hep-th]].

\bibitem{Eden:2000bk}
  B.~Eden, A.~C.~Petkou, C.~Schubert and E.~Sokatchev,
  Nucl.\ Phys.\ B {\bf 607} (2001) 191
  [hep-th/0009106].

\end{thebibliography}

\end{document}